\documentclass[aps,twocolumn,amsmath,amssymb]{revtex4}

\usepackage{graphicx}% Include figure files
\usepackage{dcolumn}% Align table columns on decimal point
\usepackage{bm}% bold math
\usepackage{amssymb}
\usepackage{amsmath}

\begin{document}

\title{Hierarchy, Fractality, Small-World and Resilience of Haversian
Bone Structure: A Complex Network Study}

\author{Luciano da Fontoura Costa}\email{luciano@if.sc.usp.br} 
\author{Matheus Palhares Viana}
    \affiliation{Instituto de F\'{\i}sica de S\~{a}o Carlos - Universidade de S\~{a}o Paulo\\
    Av. Trabalhador S\~{a}o Carlense 400, Caixa Postal 369, CEP 13560-970 \\
    S\~{a}o Carlos, S\~ao Paulo, Brazil}
\author{Marcelo Em{\'i}lio Beletti}
    \affiliation{Instituto de Ci\^{e}ncias Biom\'{e}dicas - Universidade Federal de Uberl\^{a}ndia\\
    Rua Par\'{a}, 1720, CEP 38400-902\\
    Uberl\^{a}ndia, Minas Gerais, Brazil}

\date{April 11, 2005}

\begin{abstract}

This article describes the application of recently introduced complex
networks concepts and methods to the characterization and analysis of
cortical bone structure.  Three-dimensional reconstructions of the
system of channels underlying bone structure are obtained by using
histological and computer graphics methods and then represented in
terms of complex networks.  Confluences of two or more channels are
represented as nodes, while the interconnecting channels are expressed
as edges.  The hierarchical backbone (the tree with maximum depth) of
such a network is obtained and understood to correspond to the main
structure underlying the channel system.  The remainder of the network
is shown to correspond to geographical communities, suggesting that
the bone channel structure involves a number of regular communities
appended along the hierarchical backbone.  It is shown that such
additional edges play a crucial role in enhancing the network
resilience and in reducing the shortest paths in both topology and
geometry. The recently introduced concept of fractal dimension of a
network (cond-mat/0503078) is then correlated with the resilience of
the several components of the bone channel structure to obstruction
and failure, with important implications for the understanding of the
organization and robustness of cortical bone structure.

\end{abstract}

%\pacs{Valid PACS appear here}

\maketitle

\section{Introduction}

Most biological structures and phenomena involve interactions between
several components, required in order to obtain proper functionality
and behavior.  For instance, the nervous system involves a large
number of neuronal cells interconnected through synapses, while the
skeleton consists of many bones attached one another.  Because of such
a type of organization, biological structures are often properly
represented and modeled in terms of \emph{graphs} or \emph{networks}
of interconnected nodes.  Therefore, the nervous system can be modeled
in terms of a graph whose nodes are understood to represent the
neurons and the edges the synaptic connections
(e.g.~\cite{costa05morphological}).  Interestingly, despite the
continuing interest in complex networks research
(e.g.~\cite{barabasi02statistical, newman03structure}), relatively
little attention has been focused on their use to represent
geometrical structures in biology.  One of the few exceptions is the
application of complex networks to analyze the channel network
underlying cortical bones, which was reported only
recently~\cite{viana04complex}.  The current work considerably extends
that preliminary work in several important aspects by taking into
account recently introduced powerful
concepts~\cite{costa05characterization} such as the fractal dimension
of a network~\cite{havlin05self}, the hierarchical backbones of
networks~\cite{costa04hierarchical}, as well as communities
(e.g.~\cite{newman02community}), resilience to attack
(e.g.~\cite{costa04reiforcing}) and shortest paths.  By treating such
concepts in an integrated fashion, it has been possible to infer a
series of insights about the network of channels underlying cortical
bone structures.

\section{The Haversian System}

Visual observation of the diaphysis of transversally sectioned long
bones immediately reveals that this part in the bone is formed mainly
by compact structures. However, at microscopic level, such a structure
is verified to be composed of collagen fibers organized as lamellas
with 3 to 7 $\mu m$ of thickness which are either parallel one another
or concentric to the microscopy canal, constituting the Havers System
or Osteon. Such a network, which is called the Havers channels,
follows the long axis of the bone, communicating with itself, with the
medullary cavity and with the external surface of the bone through
transversal or oblique channels called in honor of Volkmann. These
channels have as main function to contain the sanguineous vases to
nourish and supply oxygen to the deepest cells of the bone
tissue. During bone growth, its structure is constantly modified by
undoing old and forming new Havers systems. In case the distribution
of blood vessels through the bone were implemented exclusively by a
dichotomic tree, the elimination of specific channels could hinder
nutrient and oxygen supply.  Some type of architectural redundancy is
therefore expected in order to compensate for eventual channel
elimination through anastomosis during the growth or pathologies in
adults.

\section{Methodology}

The umerous of the adult cat used in this work was obtained from an
animal necropsed in the Pathology Sector of the Veterinary Hospital of
the Federal University at Uberl\^andia, Brazil.  After dissection, a
5cm thick bone ring was extracted and sliced according to histology
traditional procedures.  After digitally imaged ($700 \times 800$
pixes), the sections were registered in size, translation and
orientation, and the Havers and Volkman channels were manually
identified, in order to allow three-dimensional reconstruction of the
Haversian system by using computer graphics
techniques~\cite{foley95computer}.  A undirected graph was extracted
from the three-dimensional structure by assigning a node to each
channel confluence and an edge to each channel.  Channels shaped as a
``V'' were understood to correspond to confluence of two channels,
being therefore represented as a node with degree 2.  The complete
graph included 852 vertices and 1016 edges, implying average degree
$\left< k \right>=2.4$.

\begin{figure}
    \includegraphics[scale=0.12,clip]{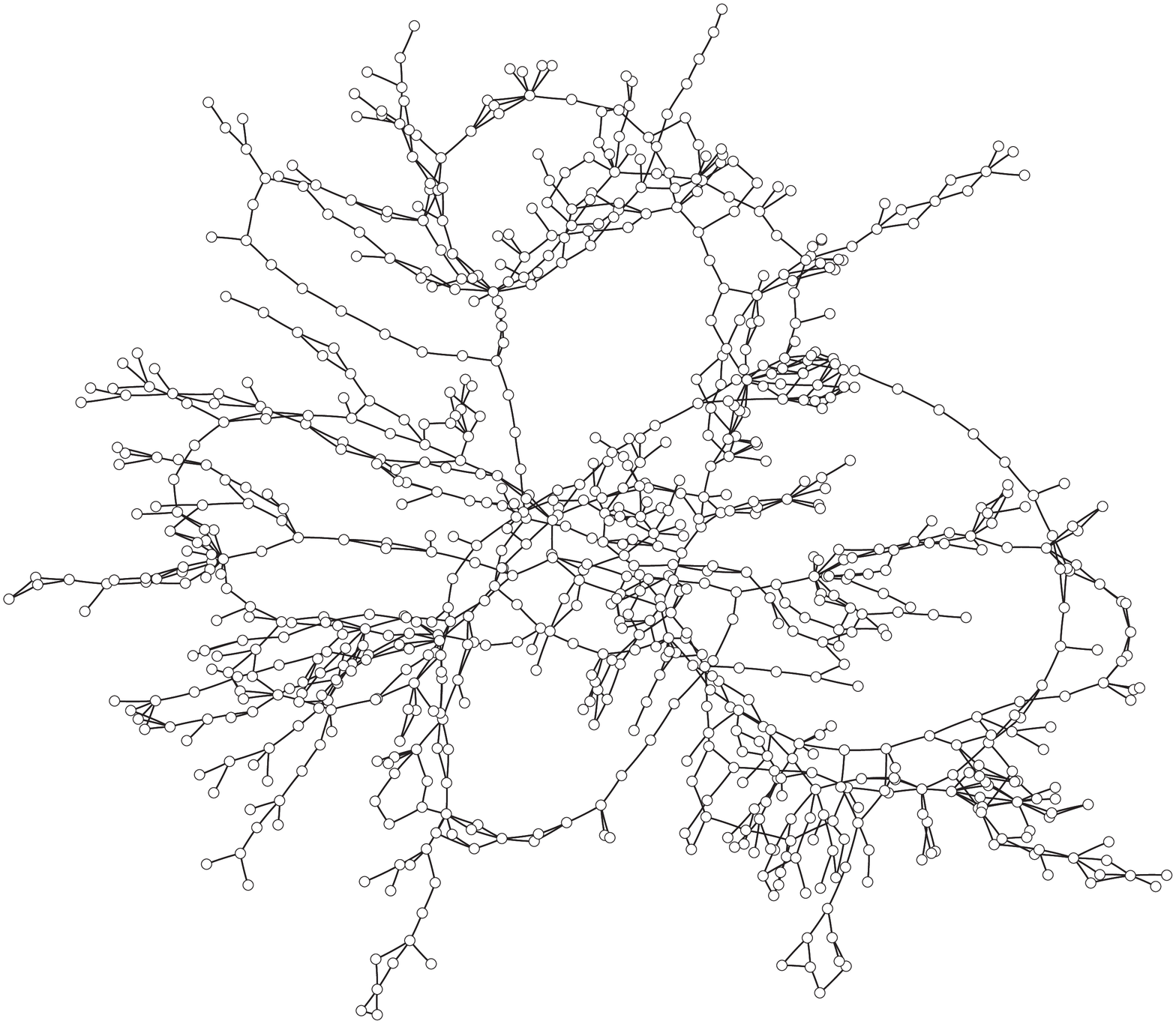} 
    \includegraphics[scale=0.12,clip]{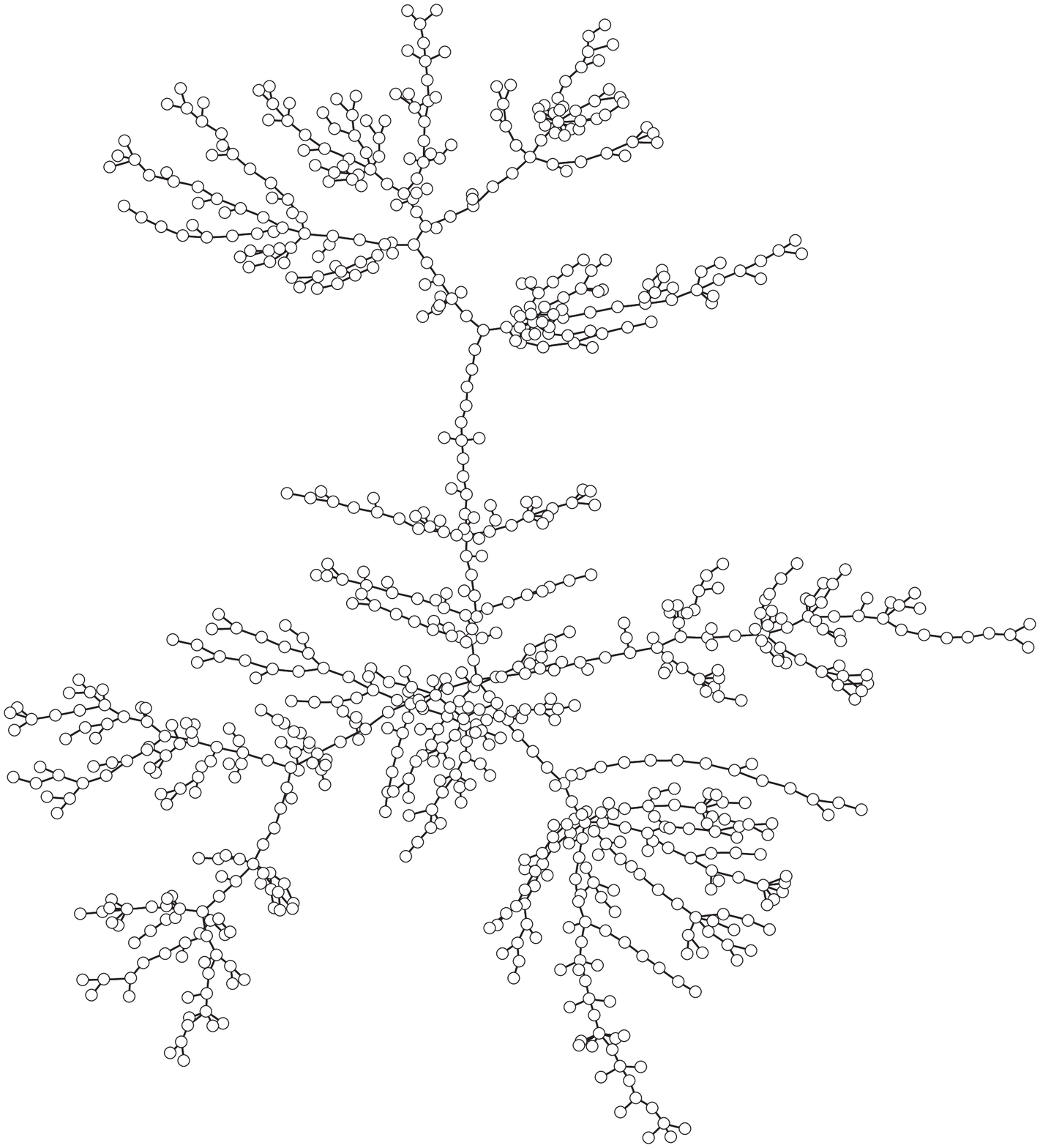}  \\

    \caption{Non geographical Haversian Network (a) and the longest
    extracted tree, understood as the respective hierarchical backbone
    (b).}
    \label{fig:tree}
\end{figure}

Figure~\ref{fig:tree}(a) shows the non-geographic representation of
the network obtained from the Haversian system.  The respective
hierarchical backbone corresponding to the longest underlying tree was
obtained as described in~\cite{costa04hierarchical} and is shown in
Figure~\ref{fig:tree}(b).  More specifically, a tree was extracted
starting at each of the network nodes.  Such trees were obtained by
following the hierarchical levels from the starting node, while
removing all cycles.  The hierarchical backbone was chosen as the
longest of such trees, but the widest tree was also identified and
considered in our investigations.  Such an approach was motivated by
the assumption that the Havers channels are meant for blood supply
along the bone structure. 

Given the spatial adjacency restrictions imposed by the 3D space where
the network lives~\cite{havlin03geographical}, it is reasonable to expect a
branched channel organization such as those found in body
vascularization and lungs.  However, by being internal to a rigid
structure (bones), such a structure would imply a serious shortcoming
in the sense that eventual obstructions of one of the channels would
mean the congestion of all downstream
channels~\cite{ross02histology}.  Therefore, we expect to find
additional channels providing bypasses along the bone channel system,
yielding a more resilient hybrid network.

\begin{figure*}
    \includegraphics[scale=0.5,clip]{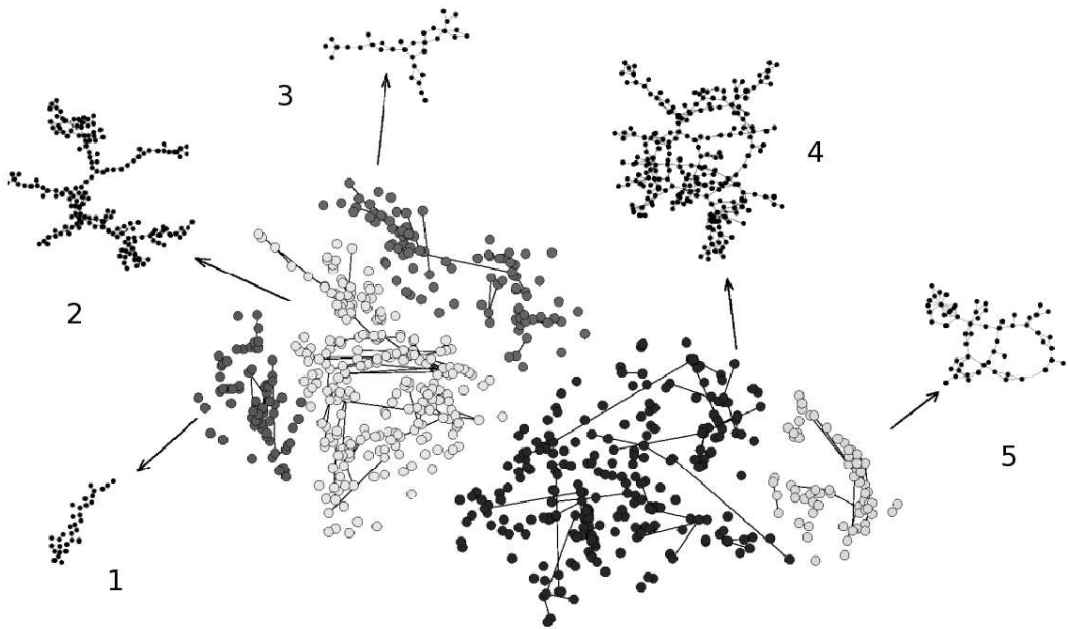}
    \caption{Haversian Network partitioned into five well-defined
    geographical communities (center) and the respectively defined
    subgraphs.}\label{partitions}
\end{figure*}

In order to verify the hypothesis above, the original channel network
was removed from the original Haversian network (by using the concept
of difference between two graphs).  The obtained results are shown in
Figure~\ref{partitions}, which includes the two-dimensional projection
of the geographic bone network (center) with its five identified
geographical communities (different gray levels), as well as the
respectively subnetworks defined by each geographical community.  Note
that the edges shown in the central structure in the previous figure
correspond to those which are part of cycles in the original
net.  Such a fact is compatible with the above hypotheses that the
Haversian system involves a tree backbone complemented by additional
bypass edges (those in the five communities), also indicating that the
latter are organized so as to form well-defined geographical
communities.  Interestingly, the same communities were obtained when
considering the widest hierarchical backbone.

\begin{figure}
    \includegraphics[scale=0.35,clip]{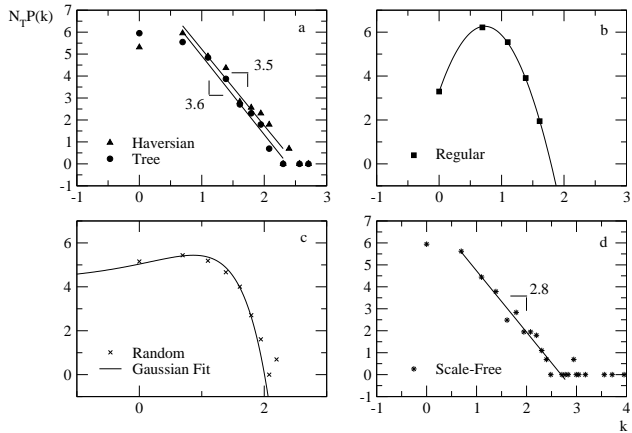}    
\caption{Degree distributions of the Haversian system and the
respective tree backbone (a).  The degree distribution of regular,
random and scale free models with the same number of nodes and
approximately the same number of edges are shown in (b-d).}
    \label{fig:degrees}
\end{figure}

The loglog distributions of the node degrees of the Haversian network
and its tree backbone are shown in Figure~\ref{fig:degrees}(a), from
which it is clear that both graphs exhibit quite similar node degrees.
The node degree distributions of regular, random and scale free
equivalent networks, i.e. with the same number of edges and nodes, are
shown in Figure~\ref{fig:degrees}(b-d), clearly illustrating that the
node degree structure of the Haversian system is not similar to any of
these alternative models, except the scale-free case in (d).

    
\begin{figure}
    \includegraphics[scale=0.35,clip]{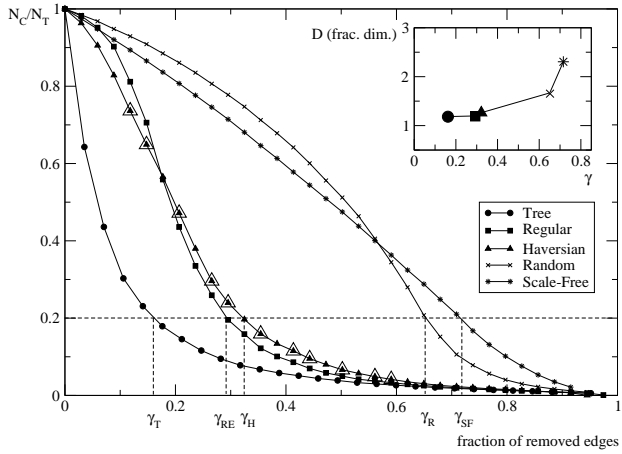}
    \caption{Despercolation of different models of complex
    networks}\label{fig:desperc}
\end{figure}

Because the resilience of the Haversian network is one of the most
important features from the functional point of view, we subjected the
original bone network to random edge attack, and the result (size of
dominant cluster) is shown in Figure~\ref{fig:desperc} (triangles) in
terms of the fraction of removed edges.  This figure also includes
results obtained for equivalent tree, regular, random and scale-free
models with approximately the same number of nodes and edges.  Note
that the Haversian network presents resilience behavior which is quite
similar to that of regular networks, suggesting that the bone
structure could have a marked degree of regularity.  However, the node
degree distributions shown in the loglog diagrams in
Figure~\ref{fig:degrees} clearly indicate that this is not the case.
In other words, the investigated Haversian channel system presents a
node degree distribution which is similar to its tree backbone (see
Figure~\ref{fig:degrees}(a)) or even a scale-free model, but a
substantially enhanced resilience which follows closely that of the
equivalent regular model.  The explanation for such mixed topological
features is provided by the fact that the Haversian system can be
understood as a hybrid, involving a tree backbone to which spatial
communities are attached in order to provide robustness to flow
obstruction, as illustrated in Figure~\ref{partitions}.

In order to obtain a more comprehensive characterization of the
topological features of the Haversian network, we calculated the
network fractal dimension as proposed by Song, Havlin and
Makse~\cite{havlin05self}, which expresses the degree of topological
self-similarity of a network.  Several interesting networks, including
protein interaction and internet, have been verified to exhibit such a
fractal organization.  The algorithm involves the coverage of the
network by boxes of size $l_B$, such that all vertices inside one of
such boxes connect one another at a distance which is smaller or equal
to $l_b-1$.  The number of boxes required to cover the whole network,
expressed as $N_B(l_B)$, is calculated for different values of $l_B$.
The topological self-similarity is characterized by the following
power law relationship

\begin{equation}
    N_B(l_B)=l_B^{-D}
    \label{frac-dim}
\end{equation}

\begin{figure}
    \includegraphics[scale=0.35,clip]{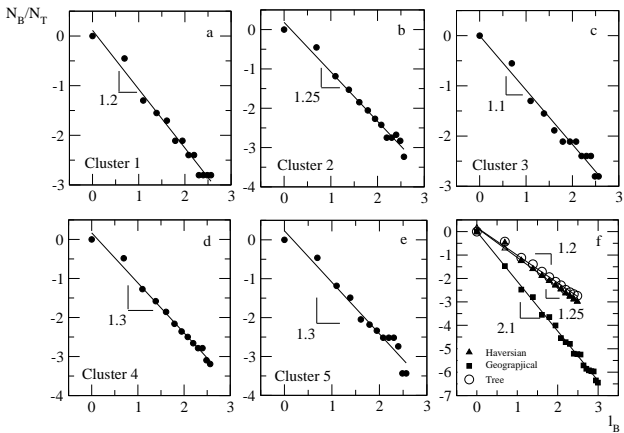}
    \caption{Fractal dimension of five communities (a-e) and of the
main Haversian network (f).  The fractal dimension of the respective
tree backbone and the box-counting fractal dimension of the
geometrical channel system are also shown in (f)} \label{fig:frac}
\end{figure}

\begin{figure}
    \includegraphics[scale=0.35,clip]{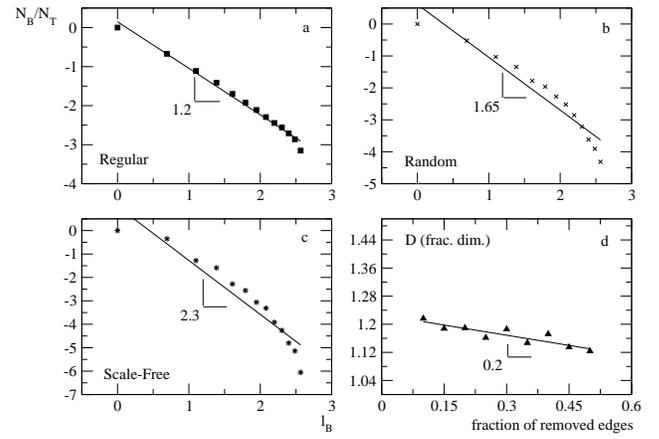}
    \caption{Fractal dimension of: (a) regular, (b) random and (c) scale-free network. In (d) shows with fractal dimension of Haversian network vary in despercolation process.} \label{fig:frac2}
\end{figure}

\begin{figure}
    \includegraphics[width=8.5cm,height=3cm,clip]{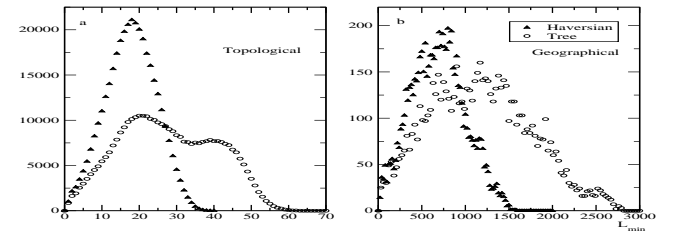}    
    \caption{Histograms of shortest topological (a) and Euclidian
distance (b) paths for the Haversian network and its respective tree
backbone.} \label{fig:short}

\end{figure}

Figures~\ref{fig:frac}(a-e) show the loglog curve of $N_B/N_T$, where
$N_T$ is the number of nodes in each case, obtained for each of the
subgraphs, while Figure~\ref{fig:frac}(f) shows the fractal dimension
for the complete Haversian network as well as its respective tree
backbone and the geometrical fractal dimension of the bone structure,
calculated by the box-counting methodology
(e.g.~\cite{russel80dimension}).  The topological fractal dimensions
of equivalent regular, random and scale free models are presented in
Figure~\ref{fig:frac2}.  The fractal nature of the regular network is
clear from (a), while the highest topological fractal dimension is
obtained for the scale free case in (c).  The topological fractal
dimension of the instances of the attacked Haversian system shown by
triangles in Figure~\ref{fig:frac2} are presented in terms of the rate
of edge attack in (d), which indicates that the topological fractal
dimension is negatively correlated with the edge attack rate,
suggesting the use of the former as an indication of the resilience of
the investigated network.  The shortest paths between any two nodes in
the Haversian network and its related tree backbone have also been
estimated and the respective histograms are shown in
Figure~\ref{fig:short}(a), while the histogram in (b) indicates the
shortest Euclidean distances.  It is clear from these results that the
addition of the communities to the tree backbone have the major effect
of reducing substantially the shortest paths.

\section{Concluding remarks}

The present work investigated the possibility of using recently
introduced complex networks concepts to represent and characterize
biological 3D structures, namely the Haversian channel system
underlying cortical bones.  The Haversian system of channels has been
found to involve a hierarchical backbone (the longest tree) to which a
series of geographically well defined communities are attached in
order to provide bypasses ensuring some degree of robustness to
eventual channel obstruction.  Such a feature of the Haversian network
has been corroborated by simulated random edge attack, which indicated
that, though exhibiting a node degree distribution similar to that of
an equivalent tree, the bone structure presents robustness close to
that of a regular network of similar size and average degree.
Additional quantification of the topological redundancy of the
Haversian system was obtained by using the recently introduced concept
of network fractal dimension, which was found to be negatively
correlated to the intensity of the attack, suggesting the use of this
measurement as an indication of the network resilience.  The presence
of the geographical communities along the tree backbone was also found
to contribute significantly to reducing the overall shortest paths in
the channel system both in topological and distance terms.

    
\begin{acknowledgments}
Luciano da F. Costa thanks HFSP RGP39/2002, FAPESP (proc. 99/12765-2)
and CNPq (proc. 3082231/03-1) for financial support.  Matheus
P. Viana is grateful to FAPESP for his MSc grant.
\end{acknowledgments}

\bibliographystyle{unsrt}
\bibliography{havers_frac}

\end{document}